\documentclass[
preprintnumbers,nofootinbib,aps,prd,floatfix]{revtex4}
\pdfoutput=1 

\usepackage{subfigure,graphicx,amsmath,amssymb,hyperref}
\usepackage{color,soul}

\newcommand{\bea}{\begin{eqnarray}}
\newcommand{\eea}{\end{eqnarray}}

\def\beq#1\eeq{\begin{align}#1\end{align}}
\def\beqnn#1\eeq{\begin{align*}#1\end{align*}}

\usepackage[usenames,dvipsnames]{xcolor}
\definecolor{darkgreen}{rgb}{0,0.5,0}

\begin{document}

\preprint{UCI-HEP-TR 2018-04}
\title{Invariant Tensors in Gauge Theories}

\author{Dillon Berger}
\email{bergerdt@uci.edu}
\author{Jessica N. Howard}
\email{jnhoward@uci.edu}
\author{Arvind Rajaraman}
\email{arajaram@uci.edu}
\affiliation{Department of Physics and Astronomy,\\
University of California, Irvine, CA 92697-4575 USA
}

\begin{abstract}
Invariant tensors play an important role in gauge theories, for example, in dualities of {$N$}=1  gauge theories. However, for theories with fields in representations larger than the fundamental,
the full set of invariant tensors { is} often difficult to construct. We present a new approach to the construction of these tensors, and use it to find the complete set of
invariant tensors of  a theory of SO(3) with fields in the symmetric tensor representation. 
\end{abstract}

\maketitle


\section{Introduction}

The invariant objects in a gauge theory play an important role. In a
gauge theory, fields transform under various representations
of the symmetry. However, the observables are gauge invariant,
and, hence, are invariant combinations of the fields.
Similarly, in a confining theory, the physical spectrum is 
composed of gauge-invariant bound states of the fundamental fields.

Similarly, duality of supersymmetric gauge theories was initiated by~\cite{Seiberg:1995pq};   
here two or more supersymmetric
gauge theories were conjectured to be equivalent
at low energies.
Testing these duality conjectures  requires detailed knowledge of the 
theory being studied. This is the case for matching the flat 
directions, matching the spectra or the chiral rings, 
or for checking 't Hooft's anomaly matching 
conditions~\cite{'tHooft:1980xb}. For all these tests, 
a knowledge of the gauge invariant operators is necessary.
Indeed, one might expect an exact equivalence of  the operators
 in these theories~\cite{Kutasov:1996ss,Brax:1999gy,Pouliot:1999yv}.

The invariant tensors of the fundamental representations of the 
classical groups are known\footnote{For a section of the mathematical 
literature relevant to invariant theory:~\cite{Weyl:1946}--\cite{Gufan:2001}.
}. However, the tensors for
the other representations are not completely classified. For
the exceptional groups such as $E_6$, it is not
even known if the full set of tensors
for the fundamental representation have been found~\cite{Pouliot:2001iw}.

Our goal in this paper is to present an approach to
finding 
the fundamental polynomial 
invariants of gauge theories  (this is sometimes known as a Hilbert basis).  
The elements of the Hilbert basis are said to be (polynomially) 
independent.
We, therefore, wish to  find the invariant tensors for
physically relevant groups and representations, with the goal
of eventually being able to classify the tensors for the various representations
of the exceptional groups. 

Our approach is to successively reduce the symmetry group
to a tractable subgroup for which the invariant structure is known.
We then find the necessary invariants in the full theory that are required
to reproduce the invariants of the subgroup. This then yields
all invariants of the larger group.

The method is more easily described by example. We begin by applying it
to two simple cases for which the answer is known. These are
(1) the group SO(2) with fields in the fundamental representation, and (2) the
group SO(3) with fields in the fundamental. The invariants of these theories
are well known, and we rederive them by our method below.

Finally, we apply our method to a nontrivial case: SO(3) with fields in the symmetric
tensor representation. The invariants here are not well known. However, we
are able to prove that there are  invariants of degree 2, 3, 4, and 5, and no others. 
We are also able to explicitly find the form of the invariants.

\section{SO(2) with fundamentals}

SO(2) is the group of 2 by 2 orthogonal matrices (we are here using the physicist approach to
groups and algebras). { A} general element can be represented as
\bea
g=\left(
\begin{array}{cc}
\cos\theta &\sin\theta
\\
-\sin\theta &\cos\theta
\end{array}\right) { ,}
\eea
where $0\ { \leq }\ \theta < 2\pi$.
The corresponding algebra is generated by the single matrix 
\bea
m=\left(
\begin{array}{cc}
0 & { -i }
\\
{ i} &0
\end{array}\right) { .}
\eea
The general group element is related to the algebra element as $g=e^{i\theta m}$.
 
 The fundamental representation can be written as a column vector of two elements
 \bea
r=\left(
\begin{array}{c}
a 
\\
b
\end{array}\right) 
\eea
(One could also impose a condition on the norm e.g. $a^2+b^2=1$, but this is not necessary).

The group acts on the representation as
\bea
r\to gr { ,}
\eea
while the infinitesimal transformations act on the representation as
\bea
r\to (1+i\theta m)r { .}
\eea

The so(2) algebra is isomorphic to a U(1) algebra. If we define $\Phi=a+ib$,
then the action of the group is just $\Phi\to \Phi e^{i\theta}$. Invariants are,
therefore, of the form $Re(\Phi_1\Phi_2)$ or $Im(\Phi_1\Phi_2)$ (or products thereof).
These can be written in SO(2) language as $r_ir'_i$ or 
$\epsilon_{ij}r_ir'_j$. 

There are thus two invariants of degree 2 in this theory. As we have seen, this is easy to show directly.
Nevertheless, we rederive this using a more generalizable method.

Accordingly, we take the approach where we try to reduce the symmetry.
We do this by choosing one particular
fundamental and giving it a fixed value
(i.e. what is called spontaneous symmetry breaking
in quantum field theory). We choose the fundamental $r^{(0)}$ and set
 \bea
r^{(0)}=\left(
\begin{array}{c}
0
\\
C
\end{array}\right)
\eea
where $C$ is a constant. We will refer to this fixed value as the vev
(short for vacuum expectation value).

This vev is fixed; therefore, the remaining symmetries 
are those that leave the vev fixed. In this case, there are no remaining symmetries.
The vev has broken the symmetry from SO(2) to nothing non-trivial.

 Now consider a second fundamental
  \bea
r^{(1)}=\left(
\begin{array}{c}
c
\\
d
\end{array}\right){ .}
\eea
Now the symmetry is completely broken; and hence the SO(2) structure
has no meaning. This is really two independent fields $c$ and $d$
with no symmetry linking them. The invariants are therefore immediate; they
are any combinations of $c$, $d$. If we add a further fundamental
  \bea
r^{(2)}=\left(
\begin{array}{c}
e
\\
f
\end{array}\right){ ,}
\eea
the invariants would be combinations of $c$, $d, e,$ and $f$.

Hence when the symmetry is broken, the invariants are
of degree 1 and are arbitrary combinations of the fields. 

What does this tell us about the original theory, assuming that we knew nothing of
the invariants of SO(2)?

We must have that the invariants of so(2) are such that after
$r^{(0)}$ gets its fixed vev,  they reduce to the degree 1
invariants of the broken theory.

The simplest way this could happen is if there were degree
1 invariants in the SO(2) theory; it is easy to check that there
are not.

The next possibility is a degree 2 invariant such as $r^{(0)}_i r^{(1)}_i$
which would become a degree 1 invariant once $r^{(0)}$ is replaced by a fixed value. Indeed,
this is { an} invariant, and becomes $Cd$ once evaluated. 
Here { $C$ } is just a constant, and is not relevant for the purposes of matching 
invariants.

We still need something that will give us the field { $c$}. This is the other degree 2
invariant $\epsilon_{ij}r^{(0)}_ir^{(1)}_j$ which become{s} $-Cc$.

These two invariants $r^{(0)}_ir^{( {1} )}_i,\epsilon_{ij}r^{(0)}_ir^{({1})}_j $ are therefore
capable of generating all invariants of the broken theory.

We can therefore assert that $r^{(a)}_ir^{(b)}_i,\epsilon_{ij}r^{(a)}_ir^{(b)}_j $ 
 generate the invariants of the full SO(2)
theory{, where $r^{(a)}_i$ and $r^{(b)}_j$ are arbitrary fundamentals} .

\section{SO(3) with fundamentals}

SO(3) is the group of 3 by 3 orthogonal matrices. 
The corresponding algebra is generated by the three  matrices
\bea
m_1=\left(
\begin{array}{ccc}
0 & { -i} &0
\\
{ i} &0&0
\\
0 &0&0
\end{array}\right)
\quad
m_2=\left(
\begin{array}{ccc}
0 &0 & { -i}
\\
0 &0&0
\\
{ i } &0&0
\end{array}\right)
\quad
m_3=\left(
\begin{array}{ccc}
0 &0 &0
\\
0 &0& \ { -i }
\\
0 & {i} &0
\end{array}\right){\  .}
\eea

 The fundamental representation can be written as a column vector of three { real} elements
 \bea
r=\left(
\begin{array}{c}
a 
\\
b
\\
c
\end{array}\right){ .}
\eea

The  infinitesimal transformations act on the representation as
\bea
r\to { \left(1+i\sum_i \alpha_i m_i\right) }r { .}
\eea

Again we try to reduce the symmetry.
We do this by choosing one particular
fundamental and giving it a fixed value.
 We choose the fundamental $r^{(0)}$ and give it a fixed vev
 \bea
r^{(0)}=\left(
\begin{array}{c}
0
\\
0
\\
C
\end{array}\right){ .}
\eea

The remaining symmetries 
are those that leave the vev fixed. In this case, this
would be a choice of $\alpha_i$ such that
\bea
r^{(0)}\to { \left(1+i\sum_i \alpha_i m_i \right) }r^{(0)}=r^{(0)}{ .}
\eea
It is fairly straightforward to go through the exercise and show that
the only possibility is $\alpha_2=\alpha_3=0$, while $\alpha_1$ is
anything. This is a residual SO(2) symmetry.
The vev has broken the symmetry from SO(3) to SO(2).

 Now consider a second fundamental
  \bea
r^{(1)}=\left(
\begin{array}{c}
d
\\
e
\\
f
\end{array}\right){ .}
\eea
Now the symmetry is  broken to SO(2). The d and e fields
form a fundamental of SO(2), while $f$ is a singlet.

There are degree 2 invariants  involving the fundamental, as described in
the previous section, as well as degree 1 invariants from the singlets.
More precisely, if we introduce
 a further fundamental
  \bea
r^{(2)}=\left(
\begin{array}{c}
g
\\
h
\\
k
\end{array}\right){ ,}
\eea
the invariants would be generated by the singlets $f, k$ 
the bilinear (dg+eh) and the bilinear (dh-gf).

We must have that the invariants of { SO}(3) are such that after
$r^{(0)}$ gets its fixed vev,  they reduce to these
invariants.

A quick check shows that we need a bilinear invariant 
$r^{(a)}_ir^{(b)}_i$ and a trilinear
invariant $\epsilon_{ijk}r^{(a)}_ir^{(b)}_jr^{(c)}_k $. (One needs to
check that these are actually invariants, but that is also
well known.) Specifically, the bilinears
$r^{(0)}_ir^{(1)}_i$, $r^{(0)}_ir^{(2)}_i$ generate the degree 1
invariants after $r^{(0)}$ gets a vev, 
the bilinear $r^{(1)}_ir^{(2)}_i$ generates the bilinear (dg+eh),
and the trilinear 
$\epsilon_{ijk}r^{(0)}_ir^{(1)}_jr^{(2)}_k $ generates 
the bilinear (dh-gf).

We can therefore assert that $r^{(a)}_ir^{(b)}_i,\epsilon_{ijk}r^{(a)}_ir^{(b)}_jr^{(c)}_j$ 
 generate the invariants of the full SO(3)
theory with fundamentals.

\section{SO(3) with symmetric tensors}

 The symmetric tensor representation can be written as a tensor $T_{ij}$
 where $T_{ij}=T_{ji}$ and $\delta^{ij}T_{ij}=0$.
 
\begin{align*}
T_{ij} =& 
\left(
\begin{array}{ccc}
 T_{11} & ~~~~~~T_{12} & ~~~T_{13} \\
 T_{12} & ~~~~~~T_{22} & ~~T_{23} \\
 T_{13} & ~~~~~~T_{23} & ~~~~-T_{11}-T_{22} \\
\end{array}
\right)\\
\end{align*}

The  infinitesimal transformations act on the representation as
\bea
T_{ij}\to T_{ij}+i(\sum_a \alpha_a m_a)_{ik}T_{kj}
+i(\sum_a \alpha_a m_a)_{jk}T_{ik} { .}
\eea

We are looking for invariants; in this case, in fact, there is no answer
in the literature for the full set of invariants. Note that we can contract the indices either using
the invariant tensor $\delta_{ij}$ or the  invariant  tensor $\epsilon_{ijk}$. However, since
 an element of the symmetric tensor representation has an even number of indices, there must be an even number of epsilon tensors,tr
which can then be converted to delta tensors. { It follows, then, that we may write a general invariant of this representation as the trace of an arbitrary (matrix) product of symmetric tensors. } We will denote such an invariant as $tr(AB\ldots N) = A_{ij}B_{kl}\ldots N_{mn} \delta_{jk}\delta_{lm}\delta_{ni}.$

Again we try to reduce the symmetry.
 We choose one fundamental $T^{(0)}$ and give it a fixed vev. Now however, there
are many possible choices which are physically
inequivalent.
We will choose 
 \bea
T^{(0)}_{11}=1\qquad T^{(0)}_{22}=1\qquad T^{(0)}_{33}=-2 { .}\nonumber
\eea
{ That is, }
\bea
T^{(0)}_{ij} =& 
\left(
\begin{array}{ccc}
 1 & 0 & ~~0 \\
0 & 1 & ~~0 \\
0 & 0 & -2 \\
\end{array}
\right){ .}
\label{Tvev}
\eea 
This breaks the SO(3) symmetry; the remaining continuous symmetry is the SO(2) symmetry
where $\alpha_1$ is nonzero, and the rest zero. 
Any other tensor ${T_{ij} }$ then decomposes 
 into 5 fields, of
charges 2, 1,0,-1,-2. The fields (with the charge labelled as a subscript) are explicitly:
\begin{align*}
 T_2~~ &= 2T_{12}+i(T_{11}-T_{22})\\
T_{-2} &= T_2^*\\
T_1 ~~&= \left(\frac{i}{\sqrt{2}}-\frac{1}{\sqrt{2}} \right)(T_{13} - iT_{23})\\
 T_{-1} &= T_1^*\\
 T_0~ ~&= T_{11} + T_{22},
\end{align*}

{ and so may be written as} 
\bea
{
T_{ij} = 
\begin{pmatrix}
 \frac{T_0}{2}+\frac{iT_{-2}}{4}-\frac{i T_2}{4} & \frac{T_{-2}}{4}+\frac{T_2}{4} & \frac{i T_{-1}}{2}-\frac{i T_1}{2} \\
 \frac{T_{-2}}{4}+\frac{T_2}{4} & \frac{T_0}{2}+\frac{T_2 i}{4}-\frac{i T_{-2}}{4} & \frac{T_{-1}}{2}+\frac{T_1}{2} \\
 \frac{i T_{-1}}{2}-\frac{i T_1}{2} & \frac{T_{-1}}{2}+\frac{T_1}{2} & -T_0 
\end{pmatrix} }
{ .}
\label{correspond} 
\eea

 In addition there is a  discrete $Z_2$ symmetry which interchanges
all indices 1 with an index of 2, so that for example $T_{13}\leftrightarrow T_{23}$.
This symmetry removes the epsilon tensor of the SO(2) symmetry.
Under the discrete  symmetry which interchanges the 1 and 2 indices, we have, 
$T_2 \leftrightarrow T_{-2}, T_1 \leftrightarrow T_{-1}, T_0 \leftrightarrow T_{0}$.

\vskip 1 cm

We now want to construct all possible combinations of these fields which
are invariant under  both the U(1) and $Z_2$  symmetries.

For the U(1), the invariants combinations are those where the charges sum to zero. These are
$T_2T'_{-2}, T_1T'_{-1}, 
T_2T'_{-1}\tilde{T}_{-1}, T_{-2}T'_{1}\tilde{T}_{1},$ and $T_0$.

However, these are not all invariant under the $Z_2$ symmetry. 
Under the symmetry, for example, $T_2T'_{-2}\to T'_2T_{-2}$, so
there are two linear combinations:
$ T_2T'_{-2}+ T'_2T_{-2}$, which is invariant under 
$Z_2$, and 
$T_2T'_{-2}- T'_2T_{-2}$, which picks up a minus
sign under 
a $Z_2$ transformation. The invariants are, then, either even or odd under the $Z_2$. 

We then find that the following expressions 
 are  invariant under
both the U(1) and the $Z_2$ symmetries
\bea
O_1&=&T_0
\\
O_2&=&Re(T_2T'_{-2})\equiv T_2T'_{-2}+ T'_2T_{-2}
\\
O_3&=&Re(T_1T'_{-1})\equiv T_1T'_{-1}+ T'_1T_{-1}
\\
O_4&=&Re(T_2T'_{-1}\tilde{T}_{-1})\equiv 
T_2T'_{-1}\tilde{T}_{-1}+T_{-2}T'_{1}\tilde{T}_{1}
\eea

On the other hand, there are combinations which are
invariant under the U(1), but odd under the $Z_2$, which are
\bea
Im(T_2T'_{-2}) &\equiv& T_2T'_{-2}- T'_2T_{-2}
\nonumber\\ 
Im(T_1T'_{-1})&\equiv&  T_1T'_{-1}- T'_1T_{-1}
\nonumber\\
Im(T_2T'_{-1}\tilde{T}_{-1})&\equiv&  T_2T'_{-1}\tilde{T}_{-1}-T_{-2}T'_{1}\tilde{T}_{1}
\nonumber
\eea
We can find
new invariants by taking pairwise products of these.
There are 6 possible pairwise products

(i) $ Im(T_1T'_{-1})Im(T_1T'_{-1})$~ ~~~~~~~~~~~~~~~~~~~~~~~~~~~~(ii) $ Im(T_1T'_{-1})Im(T_2T'_{-2})$, 

(iii) $Im(T_1T'_{-1})
(T_2T'_{-1}\tilde{T}_{-1}-T_{-2}T'_{1}\tilde{T}_{1})$
~~~~~~~~~~~~(iv) $Im(T_2T'_{-2})Im(T_2T'_{-2})$  

(v) $Im(T_2T'_{-2})
(T_2T'_{-1}\tilde{T}_{-1}-T_{-2}T'_{1}\tilde{T}_{1})$
~~~~~~~~~~~~
(vi) $(T_2T'_{-1}\tilde{T}_{-1}-T_{-2}T'_{1}\tilde{T}_{1}
)(T_2T'_{-1}\tilde{T}_{-1}-T_{-2}T'_{1}\tilde{T}_{1})$.

In principle, all these are further invariants. However, many of these can be reexpressed in 
terms of the previously found $O_{1-4}$.
In fact,
\bea
 Im(T_2T'_{-2})Im(T_2T'_{-2})
 &=&
 Re(T_2T'_{-2})
 Re(T_2T'_{-2})
 -
  Re(T_2T_{-2})
Re(T'_2T'_{-2})\nonumber
\\
 Im(T_1T'_{-1})Im(T_1T'_{-1})
 &=&
 Re(T_1T'_{-1})
Re(T_1T'_{-1})
 -
  Re(T_1T_{-1})
Re(T'_1T'_{-1})\nonumber
\\
Im(T_2T'_{-2})
Im(T_2T'_{-1}\tilde{T}_{-1})
 &=&
Re(T_2T'_{-2})
Re(T_2T'_{-1}\tilde{T}_{-1})
-
Re(T_2T_{-2})
Re(T'_2T'_{-1}\tilde{T}_{-1})\nonumber
\\
Im(T_1T'_{-1})
Im(T_2T'_{-1}\tilde{T}_{-1})
 &=&
Re(T_1T'_{-1})
Re(T_2T'_{-1}\tilde{T}_{-1})
-Re(T'_1T'_{-1})
Re(T_2T_{-1}\tilde{T}_{-1})\nonumber
\\
&&+
Re(T_1\tilde{T}_{-1})
Re(T_2T'_{-1}T'_{-1})
-
Re(T'_1\tilde{T}_{-1})
Re(T_2T_{-1}T'_{-1})
\nonumber\\
Im(T_2T'_{-1}\tilde{T}_{-1})
Im(T_2T'_{-1}\tilde{T}_{-1})
 &=&
Re(T_2T'_{-1}\tilde{T}_{-1})
Re(T_2T'_{-1}\tilde{T}_{-1})
+ {\frac{1}{2}}
Re(T_2T_{-2})
Re(T'_{-1}\tilde{T}_{1})
Re(T'_{-1}\tilde{T}_{1})
\nonumber\\
 &&- {\frac{1}{2}}
Re(T_2T_{-2})
Re(T'_{-1}\tilde{T}_{1})
Re(T'_{-1}\tilde{T}_{1})
\nonumber\\
 && -{\frac{1}{2}}
Re(T_2T_{-2})Re(T'_{-1}{T}'_{1})
Re(\tilde{T}_{-1}\tilde{T}_{1}) { .} \nonumber
\eea

However, the operator
\bea
O_5=Im(T_2T'_{-2})Im(T_1T'_{-1})\nonumber
\eea
cannot be expressed in terms of the $O_{1-4}$. 
The ring of operators  is therefore generated by the 
operators $O_{1-5}$.

 We would now like to use these operators to pick from the sea of possibilities the unique SO(3) invariant operators that allow us to generate the full space of SO(3) invariants.  So  now, the question is whether
we can 
generate all these  invariants of the broken theory 
 from SO(3) invariant 
combinations of tensors when one SO(3) tensor,
 which will
be denoted $T^{(0)}$, gets a vev
as in equation (\ref{Tvev}).  

\vskip 1 cm

We start with the simplest possibilities for
the SO(3) invariants.

No  SO(3) invariant of degree 1 exists (i.e. $tr(T)=0$
for symmetric tensors).

There is one 
SO(3) invariant of degree 2 i.e. the trace  
$tr(AB)$, where A, B are tensors.
It is convenient to
write this trace in terms of the SO(2)
fields. From
the correspondence (\ref{correspond}), we
have
\bea
tr(AB) &= (A_{-1} B_1+A_{1} B_{-1})+\frac{1}{4}(A_{-2} B_2 +A_2B_{-2})+\frac{3 A_{0} B_0}{2}
\eea

If one of these tensors  obtains
 a vev i.e. $B= T^{(0)}$, this is
 \bea
 tr(AT^{(0)})=3A_0 =3 O_1
 \eea

We 
then conclude $tr(AB)$ must be part of the 
chiral ring of SO(3), and that 
 the operator $O_1$ is generated from this element of the ring when a tensor gets a vev and breaks the group to SO(2). { That is, }
\bea
O_1= {1\over 3}tr(AT^{(0)}).
\eea

The element $tr(AB)$ also yields a SO(2)
bilinear when neither obtains a vev. This 
can be written as

\bea
 Re(A_1B_{-1})+\frac{1}{4}Re(A_2B_{-2})=tr(AB) -\frac{3 A_{0} B_0}{2}
\eea

Thus, the specific combination of bilinears $Re(A_1B_{-1})+\frac{1}{4}Re(A_2B_{-2})$ can also be generated by combinations of the bilinear element of the SO(3) chiral ring.
However, the other linear combinations of the bilinears cannot
be produced by uisng the SO(3) bilinear.
We need to use another element of the SO(3) chiral ring.

\vskip 1 cm

The next possibility is the  invariant  of degree 3: $tr(ABC)$.

Using the decomposition (\ref{correspond}), we find
\begin{align*}
tr{(ABC)}&=-\frac{A_{-1} B_{-1} C_{2}}{4}-\frac{A_{-1} B_{0} C_{1}}{4}-\frac{A_{-1} B_{1} C_{0}}{4}-\frac{A_{-1} B_{2} C_{-1}}{4}\\
&+\frac{A_{-2} B_{0} C_{2}}{8}-\frac{A_{-2} B_{1} C_{1}}{4}+\frac{A_{-2} B_{2} C_{0}}{8}\\
&-\frac{A_{0} B_{-1} C_{1}}{4}+\frac{A_{0} B_{-2} C_{2}}{8}-\frac{3 A_{0} B_{0} C_{0}}{4}-\frac{A_{0} B_{1} C_{-1}}{4}\\
&+\frac{A_{0} B_{2} C_{-2}}{8}-\frac{A_{1} B_{-1} C_{0}}{4}-\frac{A_{1} B_{-2} C_{1}}{4}-\frac{A_{1} B_{0} C_{-1}}{4}\\
&-\frac{A_{1} B_{1} C_{-2}}{4}-\frac{A_{2} B_{-1} C_{-1}}{4} +\frac{A_{2} B_{-2} C_{0}}{8}+\frac{A_{2} B_{0} C_{-2}}{8}
\end{align*}

If $C$ gets a vev, then we can set $C_0=1$ and the other $C_i$=0. This then generates the bilinear operator
\begin{align*}
tr{(ABC^{(0)})}&=-\frac{A_{-1} B_{1} }{4}
+\frac{A_{-2} B_{2} }{8}-\frac{3 A_{0} B_{0} }{4}-\frac{A_{1} B_{-1} }{4}+\frac{A_{2} B_{-2} }{8}
\end{align*}

This implies that the trilinear $tr(ABC)$ is part of the SO(3) chiral ring.
Then  the bilinear operator
$Re(A_2B_{-2})-2Re(A_1B_{-1})$ can be generated in terms of elements of
the SO(3) chiral ring. In combination with the bilinear $tr(AB)$, all operators $O_{1-3}$ can be generated.

The trilinear $tr(ABC)$ also yields a trilinear in the SO(2) fields if none of the tensors has a vev. This trilinear can be written as
\begin{align*}
tr{(ABC)}=-{1\over 4}Re(A_{-1} B_{-1} C_{2}+A_{-1} B_{2} C_{-1}+A_{2} B_{-1} C_{-1})+...
\end{align*}
where the ellipses are combinations of $O_{1-3}$. Hence the completely symmetric trilinear is
generated. To generate the other trilinears, we must consider a new SO(3) invariant.

\vskip 1 cm

The next possibility is a invariant of degree 4; $tr(ABCD)$.
However, we can narrow our search a little. The completely 
symmetrized trilinear has already been generated, and, so, we
need to look for invariants where at least two tensors are
antisymmetrized. There are exactly two such possibilities;
(a) $tr([A,B][C,D])$ and (b) $tr(ABCD)-tr(CBAD)$.
However the second can be shown to be zero for symmetric tensors.

We therefore include $tr([A,B][C,D])$ in the Hilbert basis. We 
allow the field to get a vev (i.e. 
$D_0=1$ and the other $D_i$=0.). We then find
\bea
tr([A,B][C,D^{(0)}])=\frac{3}{2}(Re(A_{-1}C_{-1}B_2) -Re(B_{-1}C_{-1}A_2))+...
\eea
where the ellipses are combinations of $O_{1-3}$.
We therefore generate the trilinear
operator $(Re(A_{-1}C_{-1}B_2) -Re(B_{-1}C_{-1}A_2))$.

We also find that the $tr([A,C][B,D])$ generates
$(Re(A_{-1}B_{-1}C_2) -Re(C_{-1}B_{-1}A_2))$.

Along with the completely symmetric trilinear, we have generated all trilinears 
from $O_4$.
This implies that the operator $tr([A,B][C,D])$ is part of the SO(3) chiral ring.

\vskip 1 cm
We now turn to the generation of all terms of the form $O_5$. It is useful, however,
to first work out what structures we are looking for. The $O_5$ terms
 have the general form $Im(A_1B_{-1})Im(C_2D_{-2})$. There are other 
such forms related by a permutation acting on ABCD. There are 6 such
independent permutations which are 
 $Im(A_1B_{-1})Im(C_2D_{-2})$,  $Im(A_1C_{-1})Im(B_2D_{-2})$, etc.

It is useful to
 separate these  into combinations which are (anti)symmetric when we interchange the
two pairs of fields. In the symmetric terms
we further isolate the expression which is antisymmetric under any
pairwise exchange. These are

\bea
Im(A_1B_{-1})Im(C_2D_{-2})\pm Im(C_1D_{-1})Im(A_2B_{-2})\nonumber
\\
Im(A_1C_{-1})Im(B_2D_{-2})\pm Im(B_1D_{-1})Im(A_2C_{-2})\nonumber
\\
Im(A_1D_{-1})Im(B_2C_{-2})\pm Im(B_1C_{-1})Im(A_2D_{-2})\nonumber
\eea

If no field gets a vev, the operator $tr([A,B][C,D])$ generates a operator of the
form $O_5$:
\bea
tr([A,B][C,D])=
 -(Im(A_1B_{-1})Im(C_2D_{-2}) +Im(A_2B_{-2})Im(C_1D_{-1}))\\
+\frac{1}{2}(Im(A_1D_{-1})Im(B_2C_{-2}) + Im(A_2D_{-2})Im(B_1C_{-1}))\\
-\frac{1}{2} (Im(A_1C_{-1})Im(B_2D_{-2})+Im(A_2C_{-2})Im( B_1D_{-1}) ) + ...\\
\eea

Similarly, we also have
\bea
tr([A,C][B,D])=
 -(Im(A_1C_{-1})Im(B_2D_{-2}) +Im(A_2C_{-2})Im(B_1D_{-1}))\\
-\frac{1}{2}(Im(A_1D_{-1})Im(B_2C_{-2}) + Im(A_2D_{-2})Im(B_1C_{-1}))\\
-\frac{1}{2} (Im(A_1B_{-1})Im(C_2D_{-2})+Im(A_2B_{-2})Im( C_1D_{-1}) ) + ...\\
\eea

Since we have 
\bea
tr([A,B][C,D])-tr([A,C][B,D])-tr([A,D][C,B])=0
\eea
the completely antisymmetrized structure cannot be generated.
The operators $tr([A,B][C,D]), tr([A,C][B,D])$ thus generate 
two combinations of the $O_5$ form
which are symmetric under interchange of (AB), (CD), but not
the fully antisymmetric combination.
We therefore still need new SO(3) invariants.

\vskip 1 cm 
We must consider invariants of degree 5. These are of the form
$tr(ABCDE)$ where one field gets a vev, since we must generate a quadrilinear. 
We can take $E$ to be the field getting a vev; there are then 24 possible
permutations of the remaining field in this invariant. 

We first try to generate the completely antisymmetric (in ABCD) structure. It
is relatively easy to see that
this requires the SO(3) chiral ring
to include the structure tr(ABCDE) where all five tensors are completely antisymmetrized.

We then generate the structures in which (AB) (CD) are antisymmetric, and
where the structure is odd under pairwise interchange of (AB), (CD). The 
only possible nonzero
candidate
structure is
\bea
tr(A[C,D]BE)
-tr(C[A,B]DE)
\eea
We therefore 
include the last candidate as part of the chiral ring,
and find its decomposition
once $E$ gets a vev. This is
\begin{align*}
tr(BE^{(0)}A[C,D] -DE^{(0)}C[A,B])&=\frac{3}{2}(Im(C_1D_{-1})Im(A_{2}B_{-2}) -
Im(C_2D_{-2})Im(A_1B_{-1}) ) \\
tr(CE^{(0)}B[D,A] -AE^{(0)}D[B,C])&=\frac{3}{2}( Im(A_1D_{-1})Im(B_{-2}C_{2}) -Im(A_2D_{-2})Im(B_{-1}C_1) )  \\
tr(CE^{(0)}A[D,B]-BE^{(0)}D[A,C]) &= \frac{3}{2}Im(A_{2}C_{-2})Im(B_{-1}D_{1}) - 
Im(A_{1}C_{-1}) )Im(B_{-2}D_{2}) 
\end{align*}
which, indeed, generates all the required $O_5$ structures.

In summary, we have found the Hilbert basis for the SO(3) group with symmetric tensors. These are:
\bea
S_1&=&tr(AB)
\\
S_2&=&tr(ABC)
\\
S_3&=&tr([A,B][C,D])
\\
S_4&=&tr(ABCDE)\quad  {\rm completely\ antisymmetrized\ in\ ABCDE}
\\
S_5&=&tr(A[C,D]BE)-tr(C[A,B]DE)
\eea

\section{Conclusions}

In this note, we have presented a new approach to constructing the invariant tensors for theories
with arbitrary representation content. We have applied this method to simple examples and shown
that the known results are reproduced. We have also applied this method to a  theory where the
chiral ring was not known -
a SO(3) theory with symmetric tensors - and shown that we can extract the chiral ring
by our methods.

While we have worked  with
relatively simple groups here we expect that our methods
can be generalized straightforwardly to other groups and representations. 
In particular, it would be interesting to find invariant tensors for SU({$N$})
tensors with adjoint representations, which are interesting for the physics 
of strong interactions, and
for exceptional groups, which are important for applications to dualities in supersymmetric theories.

\section*{Acknowledgements}

This work was supported in part by NSF Grant No.~PHY-1620638.


\begin{thebibliography}{19}


\bibitem{Seiberg:1995pq}
N.~Seiberg,
``Electric - magnetic duality in supersymmetric nonAbelian gauge theories,''
Nucl.\ Phys.\ B {\bf 435}, 129 (1995)
[hep-th/9411149].



\bibitem{'tHooft:1980xb}
G.~.~'t Hooft, C.~.~Itzykson, A.~.~Jaffe, H.~.~Lehmann, 
P.~K.~Mitter, I.~M.~Singer and R.~.~Stora,
``Recent Developments In Gauge Theories. Proceedings, 
Nato Advanced Study Institute, Cargese, France, August 26 - 
September 8, 1979,''
{\it  New York, Usa: Plenum (1980) 438 P. (Nato Advanced 
Study Institutes Series: Series B, Physics, 59)}.



\bibitem{Kutasov:1996ss}
D.~Kutasov, A.~Schwimmer and N.~Seiberg,
``Chiral Rings, Singularity Theory and Electric-Magnetic Duality,''
Nucl.\ Phys.\ B {\bf 459}, 455 (1996)
[hep-th/9510222].



\bibitem{Brax:1999gy}
P.~Brax, C.~Grojean and C.~A.~Savoy,
``Anomaly matching and syzygies in N = 1 gauge theories,''
Nucl.\ Phys.\ B {\bf 561}, 77 (1999)
[hep-ph/9808345].

\bibitem{Pouliot:1999yv}
P.~Pouliot,
``Molien function for duality,''
JHEP {\bf 9901}, 021 (1999)
[hep-th/9812015].



\bibitem{Weyl:1946}
H.~Weyl, ``The Classical Groups, Their Invariants and 
Representations,'' Princeton University Press, 1946.

\bibitem{Fulton:1991}
W.~Fulton and J.~Harris, ``Representation Theory, A First Course,'' 
Springer-Verlag, 1991.

\bibitem{Olver:1999}
P.~Olver, ``Classical Invariant Theory,'' London Mathematical 
Society Student Texts~\#44, 1999.




\bibitem{Dieudonne:1971}
J.~Dieudonne and J.~Carrell, ``Invariant Theory, Old and New,'' 
Academic Press, 1971.

\bibitem{Howe:1995}
R.~Howe, ``Perspectives on Invariant Theory,'' The Schur 
Lectures (1992), Israel Mathematical
Conference Proceedings, 1995.

\bibitem{Gurevich:1983}
G.~Gurevich, ``Foundations of the Theory of Algebraic Invariants,'' 
P. Noordhoff - Groningen, The Netherlands, 1964.

\bibitem{schwarz:1983}
G.~Schwarz, 
``Invariant theory of $G_2$,''
Bull.\ Am.\ Math.\ Soc. {\bf 9}, 335 (1983).

\bibitem{Schwarz:1988}
G.~Schwarz,
``Invariant theory of $G_2$ and $Spin_7$,''
Comment.\ Math.\ Helvetici {\bf 63}, 624 (1988).

\bibitem{Schwarz:1978a}
G.~Schwarz,
``Representations of Simple Lie Groups with a Free Module of Covariants,''
Inventiones Math.\ {\bf 50}, 1 (1978).

\bibitem{Schwarz:1978b}
G.~Schwarz,
``Representations of Simple Lie Groups with Regular Rings of Invariants,''
Inventiones Math.\ {\bf 49}, 167 (1978).

\bibitem{Popov:1984}
V.~Popov,
``Syzygies in the Theory of Invariants,''
Math.\ USSR\ Izvestiya {\bf 22}, 507 (1984).

\bibitem{Gufan:2001}
Y.~Gufan, Al.~V.~ Popov, G.~Sartori, V.~Talamini, G.~Valente and E.~Vinberg,
``Geometric Invariant Theory Approach to the Determination of 
Ground States D-wave 
Condensates in Isotropic Space,'' 
Jour.\ Math.\ Phys.\ {\bf 42}, 1533 (2001).



	
	
	
	
	
	






\bibitem{Pouliot:2001iw} 
  P.~Pouliot,
  J.\ Phys.\ A {\bf 34}, 8631 (2001)
  doi:10.1088/0305-4470/34/41/317
  [hep-th/0107151].
	
	

\bibitem{Buccella:1982nx}
F.~Buccella, J.~P.~Derendinger, S.~Ferrara and C.~A.~Savoy,
``Patterns Of Symmetry Breaking In Supersymmetric Gauge Theories,''
Phys.\ Lett.\ B {\bf 115}, 375 (1982).

\bibitem{Procesi:1985hr}
C.~Procesi and G.~W.~Schwarz,
``The Geometry Of Orbit Spaces And Gauge Symmetry Breaking In 
Supersymmetric Gauge Theories,''
Phys.\ Lett.\ B {\bf 161} (1985) 117.

\bibitem{Gatto:1987bt}
R.~Gatto and G.~Sartori,
``Consequences Of The Complex Character Of The Internal Symmetry In 
Supersymmetric Theories,''
Commun.\ Math.\ Phys.\  {\bf 109}, 327 (1987).

\bibitem{Luty:1996sd}
M.~A.~Luty and W.~I.~Taylor,
``Varieties of vacua in classical supersymmetric gauge theories,''
Phys.\ Rev.\ D {\bf 53}, 3399 (1996)
[hep-th/9506098].

\bibitem{Gherghetta:1996dv}
T.~Gherghetta, C.~Kolda and S.~P.~Martin,
``Flat directions in the scalar potential of the supersymmetric 
standard model,''
Nucl.\ Phys.\ B {\bf 468}, 37 (1996)
[hep-ph/9510370].

\bibitem{Brax:2001an}
P.~Brax and C.~A.~Savoy,
``Supersymmetric flat directions and analytic gauge invariants,''
hep-th/0104077.

\bibitem{Csaki:1998aw}
C.~Csaki and H.~Murayama,
``Discrete anomaly matching,''
Nucl.\ Phys.\ B {\bf 515}, 114 (1998)
[hep-th/9710105].

\bibitem{Argyres:1996eh}
P.~C.~Argyres, M.~Ronen Plesser and N.~Seiberg,
``The Moduli Space of N=2 SUSY {QCD} and Duality in N=1 SUSY {QCD},''
Nucl.\ Phys.\ B {\bf 471}, 159 (1996)
[hep-th/9603042].

\bibitem{Berkooz:1996cb}
M.~Berkooz,
``A Comment on Non-Chiral Operators in S{QCD} and its Dual,''
Nucl.\ Phys.\ B {\bf 466}, 75 (1996)
[hep-th/9512024].

\bibitem{LiE:1996}
M.~van Leeuwen, A.~Cohen and B.~Lisser, ``LiE Manual,'' version 2.1,
Amsterdam, The Netherlands, 1996.

\end{thebibliography}
\end{document}